\documentclass[runningheads]{llncs}
\usepackage[T1]{fontenc}
\usepackage{graphicx}
\begin{document}
\title{Carbon Per Transistor (CPT): The Golden Formula for Green Computing Metrics}
\titlerunning{Carbon Per Transistor (CPT)}
%
\author{Zag ElSayed\inst{1} \and
Nelly Elsayed\inst{2} \and
Ahmed Abdelgawad\inst{3}}
\authorrunning{Z. ElSayed et al.}
%
\institute{University of Cincinnati, Ohio, USA \and
School of Information Technology, University of Cincinnati, Ohio, USA \and
School of Eng. \& Tech. Central Michigan University, Michigan, USA
}
\maketitle          
\begin{abstract}
As computing power advances, the environmental cost of semiconductor manufacturing and operation has become a critical concern. However, current sustainability metrics fail to quantify carbon emissions at the transistor level, the fundamental building block of modern processors. This paper introduces a Carbon Per Transistor (CPT) formula, a novel approach and green implementation metric to measuring the $CO_2$ footprint of semiconductor chips from fabrication to end-of-life. By integrating emissions from silicon crystal growth, wafer production, chip manufacturing, and operational power dissipation, the CPT formula provides a scientifically rigorous benchmark for evaluating the sustainability of computing hardware. Using real-world data from Intel Core i9-13900K, AMD Ryzen 9 7950X, and Apple M1/M2/M3 processors, we reveal a startling insight—manufacturing emissions dominate, contributing 60–125 $kg$ $CO_2$ per CPU, far exceeding operational emissions over a typical device lifespan. Notably, Apple’s high-transistor-count M-series chips, despite their energy efficiency, exhibit a significantly larger carbon footprint than traditional processors due to extensive fabrication impact. This research establishes a critical reference point for green computing initiatives, enabling industry leaders and researchers to make data-driven decisions in reducing semiconductor-related emissions and get correct estimates for the green factor of the information technology process. The proposed formula paves the way for carbon-aware chip design, regulatory standards, and future innovations in sustainable computing.

\end{abstract}
\section{Introduction}
\begin{figure}[htbp]
\centerline{\includegraphics[width =\linewidth]{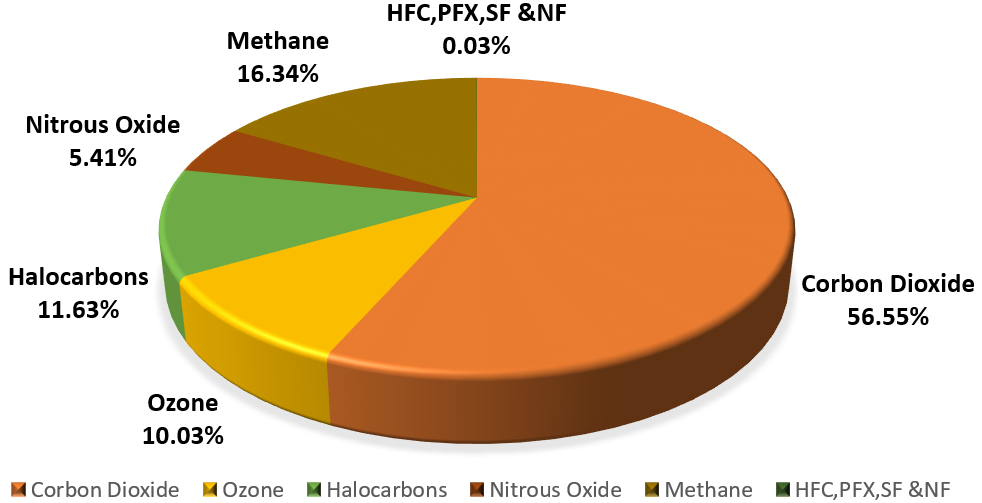}}
\caption{The proposed CryptoDNA model architecture.}
\label{fig1}
\end{figure}

The rapid advancement of semiconductor technology has led to an exponential increase in computational power, enabling breakthroughs in artificial intelligence, high-performance computing, and mobile communications. However, this progress comes at a significant environmental cost\cite{ma2024industry}. The fabrication of modern high-density processors demands substantial energy inputs, specialized chemicals, and complex multi-stage processes that release large quantities of $CO_2$ and other greenhouse gases (GHGs) reported by \cite{kuo2022assessing} and the US Environmental Protection Agency (EPA 2025), shown in Fig\ref{fig1}.

To address this, we propose a Carbon Per Transistor (CPT) formula, a mathematical model that systematically quantifies the $CO_2$ footprint associated with semiconductor fabrication and operation. By applying this model to Intel, AMD, and Apple processors, we provide a benchmarking tool for green computing, allowing policymakers, manufacturers, and researchers to design low-carbon computing architectures.

\subsection{The Growing Carbon Footprint of Computing}
The exponential increase in transistor density, as predicted by Moore’s Law\cite{moore1998cramming}, has enabled high-performance computing while simultaneously escalating its environmental impact. The semiconductor industry accounts for a significant fraction of global carbon emissions, primarily due to the energy-intensive processes involved in chip fabrication and the subsequent power dissipation during operation\cite{boyd2011life}.

Traditional carbon footprint assessments of computing systems focus on the total energy consumption of data centers or end-user devices without granular analysis at the transistor level. Given that modern processors contain upwards of 10–100 billion transistors, quantifying emissions per transistor is crucial for establishing sustainable semiconductor design principles.

This work proposes a Carbon Per Transistor (CPT) formula, integrating emissions from:

\begin{equation}
C_{total} = C_{crystal} + C_{wafer} + C_{man} + C_{oper}\label{eq1}
\end{equation}

where $C_{crystal}$ accounts for crystal growth and silicon purification, $C_{wafer}$ represents wafer slicing and doping, $C_{man}$ manufacturing covers photolithography, etching, and chip packaging, and $C_{op}$ corresponds to transistor power dissipation over its lifetime.

\subsection{Semiconductor Manufacturing and $CO_2$ Emissions}
Semiconductor fabrication requires ultra-pure silicon, processed through an energy-intensive Czochralski process, emitting an estimated 40 kg $CO_2$ per 300mm wafer\cite{wang2023environmental}. Wafer processing, involving chemical vapor deposition (CVD), ion implantation, and plasma etching, further contributes to emissions, with each transistor fabrication cycle emitting approximately 2–5 $\mu g$ $CO_2$\cite{liu2024sustainable}. For a given processor with $N$ transistors, the total manufacturing $CO_2$ footprint can be approximated as:

\begin{equation}
C_{man} = (N \times C_{trans})_{fab}\label{eq2}
\end{equation}

where $C_{trans}$ is the pre-transistor fabrication emission factor. Beyond manufacturing, a transistor’s lifetime power dissipation contributes significantly to its cumulative $CO_2$ footprint. Assuming an average power per transistor $P_{trans}$ ($W$) and an operational duration of $H_{lifetime}$ ($hours$), the operational $CO_2$ emission ($C_{opr}$) is calculated as:

\begin{equation}
C_{oper} = (P_{trans} \times H_{lifetime} \times EF)\label{eq3}
\end{equation}
where $EF$ is the grid emission factor ($kg$ $CO_2$ per $kWh$), $H_{lifetime}$ represents the device lifespan in hours. Regarding thermal design power (TDP) and as a benchmark we analyze real-world processors we analysed are shown in Table\ref{tabel1}

\begin{table}[htbp]
\caption{The benchmark real-world processors specifications }
\begin{center}
\begin{tabular}{|c|c|c|c|}
\hline
\textbf{ID}&\textbf{Processor}&\textbf{\# Transistors}&\textbf{Power TDP}\\
\cline{1-4}
\hline
1&Intel Core i9-13900K&~12 billion&125–253W   \\
\hline
2&AMD Ryzen 9 7950X&~13.14 billion&170–230W   \\
\hline
3&Apple M1, M2, and M3&~16–25 billion&20–22W  \\
\hline
\end{tabular}
\label{tabel1}
\end{center}
\end{table}

Through empirical analysis, we demonstrate that manufacturing emissions outweigh operational $CO_2$ contributions, making chip fabrication the dominant factor in semiconductor sustainability.

\subsection{Contribution of This Work}
The key contributions of this work are the development of the CPT formula, enabling precise $CO_2$ estimation at the transistor level. Quantification of manufacturing per operational emissions, highlighting the dominance of fabrication processes. Performe empirical validation and verification of the proposed formula using commercial processors, presenting a comparative analysis of Intel, AMD, and Apple silicon chips. Moreover, providing a reference framework for green computing strategies, aiding semiconductor manufacturers in low-carbon design choices. By integrating engineering principles, mathematical modeling, and empirical data, this study establishes a scientific benchmark for evaluating semiconductor sustainability, paving the way for future advancements in carbon-efficient chip design.

\section{Related Work}
\subsection{Carbon Footprint of Semiconductor Manufacturing}
The environmental impact of semiconductor fabrication has been extensively studied, with a primary focus on energy consumption and greenhouse gas (GHG) emissions. The studies in\cite{huang2016developing} and \cite{vasan2014carbon} conducted a life cycle assessment (LCA) of semiconductor manufacturing, revealing that the wafer fabrication stage alone contributes up to 80\% of total emissions due to the high-energy demands of ion implantation, photolithography, and chemical vapor deposition (CVD). The study highlighted that perfluorinated compounds (PFCs), a class of synthetic compounds containing thousands of chemicals formed from carbon chains with fluorine attached to these chains, used in plasma etching and chamber cleaning have global warming potentials (GWP) thousands of times higher than $CO_2$, making their reduction a critical challenge.

Similarly, the studied in \cite{huang2016developing} and \cite{boyd2011life} analyzed emissions from 300$mm$ wafer fabrication, estimating that each 5$nm$ process node chip requires over 450 $kWh$ of energy per wafer, translating to approximately 35–50 $kg$ $CO_2$ per wafer under a global average electricity grid mix. The study suggested moving to renewable energy sources in semiconductor foundries as an effective strategy to reduce emissions.

However, these studies provide macro-scale insights into wafer-level carbon footprints but lack transistor-level granularity. The per-transistor $CO_2$ footprint remains poorly quantified, limiting the precision of green computing benchmarks. Our work addresses this critical gap by introducing a mathematical framework to compute Carbon Per Transistor (CPT) across different processors.

\subsection{Power Dissipation and Energy Efficiency in Computing}
The operational energy consumption of transistors is a key factor in sustainable computing. The legendary work in \cite{dennard1974design} originally predicted that as transistor size shrinks, power consumption per transistor should decrease proportionally. However, due to power density limitations, modern processors face the end of Dennard Scaling, leading to increased dynamic and leakage power dissipation\cite{horowitz20141}.
Several studies have examined power efficiency trends in computing. The work in\cite{koomey2011growth} demonstrated that the energy efficiency of computing devices doubles approximately every 1.5 years, following an exponential improvement trend. The authors in\cite{hennessy2011computer} emphasized that while performance-per-watt has improved due to better architecture and process scaling, the total energy consumption of modern chips remains high due to the increasing transistor count in multi-core architectures. Moreover, the work in \cite{sungheetha2024adaptive} proposed chip-level optimizations, such as dynamic voltage scaling (DVS), near-threshold computing, and specialized AI accelerators, to reduce operational power consumption. Despite these efforts, no existing research explicitly quantifies $CO_2$ emissions at the transistor level. This study bridges that gap by combining power dissipation analysis with semiconductor manufacturing emissions, establishing a rigorous $CO_2$ footprint per transistor metric.

\subsection{Semiconductor Industry Efforts Toward Green Computing}
Leading semiconductor manufacturers have initiated efforts to reduce their carbon footprint. Taiwan Semiconductor Manufacturing Company Limited (TSMC) has committed to achieving net-zero emissions by 2050, with initiatives including 100\% renewable energy adoption and process efficiency improvements \cite{nagapurkar2023cleaner}. Additionally, Intel has pledged to achieve zero carbon fabs by 2040, focusing on carbon-neutral chip packaging and low-emission materials\cite{ivan2021practices}. Furthermore, AMD, and NVIDIA have introduced low-power AI accelerators, optimizing performance-per-watt to reduce the energy impact of data center AI workloads\cite{kang2024sustainability}. While these initiatives are commendable, they lack a standardized, quantifiable metric to assess the sustainability of individual transistor designs. The Carbon Per Transistor (CPT) formula proposed in this study provides such a metric, enabling comparative sustainability analysis across different architectures, process nodes, and device categories.

The related works reviewed above establish the need for a scientific metric quantifying per-transistor carbon emissions. Existing research has examined wafer-level emissions and energy use. Still, it lacks transistor-level granularity, processor power dissipation trends but does not compute $CO_2$ emissions per transistor, and Industry-led sustainability efforts but lacks a standardized $CO_2$ metric for green computing.

However, to our knowledge, no proposed Carbon Per Transistor (CPT) formula to calculate $CO_2$ emissions per transistor across manufacturing and operation was done or validated, and no established benchmark for sustainable semiconductor design exists yet. By bridging these research gaps, this work advances the state-of-the-art in carbon-conscious computing, paving the way for low-carbon semiconductor design and green computing metrics.

\section{Formula Development}
As semiconductor technology advances toward denser, more power-efficient architectures, the carbon footprint of individual transistors becomes a fundamental metric for evaluating sustainability. While previous research has quantified wafer-level and processor-level $CO_2$ emissions, no standardized equation exists for determining the precise $CO_2$ impact per transistor.

This section introduces the Carbon Per Transistor (CPT) formula, a groundbreaking approach to quantifying the total carbon footprint of a single transistor, integrating manufacturing emissions (fabrication, lithography, packaging) and operational emissions (power dissipation, grid carbon intensity). This equation serves as the cornerstone of green computing analytics, allowing engineers, policymakers, and industry leaders to assess the sustainability of semiconductor devices with mathematical constraints.

\subsection{The General Carbon Per Transistor (CPT) Formula}

We define the total $CO_2$ footprint per transistor as the sum of its manufacturing emissions and operational emissions:

\begin{equation}
C_{trans} = C_{man} + C_{oper}\label{eq4}
\end{equation}

where: $C_{man}$ is the $CO_2$ emitted per transistor during fabrication ($\mu g$ $CO_2$/transistor). $C_{oper}$ is the $CO_2$ emitted per transistor over its entire operational lifetime ($mg$ $CO_2$/transistor).
The total carbon footprint for an entire processor (CPU/GPU/SoC) containing $N_{trans}$ transistors is:

\begin{equation}
C_{total} = N_{trans} \times C_{trans}\label{eq5}
\end{equation}

This formula directly compares processors from different architectures, semiconductor nodes, and power profiles, making it a universal metric for sustainable chip design.

\subsection{Manufacturing Emissions Per Transistor}
Semiconductor fabrication is an extremely energy-intensive process involving wafer growth, photolithography, etching, ion implantation, chemical vapor deposition (CVD), and packaging. We define the per-transistor manufacturing $CO_2$ emissions as:

\begin{equation}
C_{man} = \frac {C_{wafer}}{\Upsilon \times  N_{trans/wafer} }\label{eq6}
\end{equation}

where $C_{wafer}$ is the total $CO_2$ emissions per wafer ($kg$ $CO_2$/wafer), $\Upsilon$ is the wafer yield factor, and $N_{trans/wafer}$ is the total transistors per wafer.

Empirical Estimation for Leading Process Nodes: for a 300$mm$ wafer, industry reports indicate:
\begin{itemize}
    \item ~450 $kg$ $CO_2$ per wafer for 5$nm$ technology.
    \item ~350 $kg$ $CO_2$ per wafer for 7$nm$ technology.
    \item ~350 $kg$ $CO_2$ per wafer for 7$nm$ technology.
\end{itemize}

With modern wafers containing hundreds of billions of transistors, we derive estimated manufacturing $CO_2$ per transistor:
\begin{equation}
C_{man} \approx 2 - 5 \mu g CO_2/transistor \label{eq7}
\end{equation}

Note: for sub-7nm nodes, fabrication is the dominant source of emissions.

\subsection{Operational $CO_2$ Emissions Per Transistor}
The operational carbon footprint of a transistor depends on its power dissipation and usage duration. The total $CO_2$ emissions from power consumption are given by:

\begin{equation}
C_{opr} = P_{trans} \times H_{lifetime} \times EF \label{eq8}
\end{equation}

where $P_{trans}$ is the average power consumption per transistor ($W$), $H_{lifetime}$ is the total lifetime usage in hours. $EF$ is the carbon emission factor ($kg$ $CO_2$ per kWh). For a given processor (CPU/GPU/SoC), the per-transistor power consumption is estimated as:

\begin{equation}
P_{trans}  = \frac{P_{total}}{N_{trans}} \label{eq9}
\end{equation}
where $P_{total}$ is the total processor power consumption ($W$).

Empirical Estimation for Modern Chips: Using real-world processor data:
\begin{itemize}
    \item Intel Core i9-13900K $\to 10.4 - 21.1 nW$ per transistor.
    \item AMD Ryzen 9 7950X $\to 12.9 - 17.5 nW$ per transistor.
    \item Apple M3 $\to 0.8 - 0.8 nW$ per transistor.
\end{itemize}

Assuming: five (5) years of daily use (8 hours/day) and a global energy grid emission factor of 0.4 $kg$ $CO_2/kWh$, we estimate:

\begin{equation}
C_{oper} \approx 60 - 250 mg  CO_2/transistor \label{eq10}
\end{equation}
Note: indicating that operational emissions are significant but secondary to manufacturing emissions.

\section{The full CPT Formula}

By combining the manufacturing and operational components, we arrive at the full CPT formula:

\begin{equation}
C_{trans}  = \frac{C_{wafer}}{\Upsilon \times N_{trans/wafer}} +P_{trans} \times H_{lifetime} \times EF \label{eq11}
\end{equation}

This proposed novel equation provides a scientific foundation for evaluating semiconductor sustainability, enabling at innovation at least (but not limited to) three (3) categories:
\begin{itemize}
    \item Industry benchmarking: Manufacturers can optimize process nodes for lower manufacturing emissions.
    \item Green computing policies: Policymakers can enforce $CO_2$ reduction standards based on chip efficiency.
    \item Sustainable design innovations: Engineers can develop carbon-aware architectures, reducing both fabrication and operational footprint.
\end{itemize}

Over and above that, we summarize formulas as a granular and insightful view of semiconductor sustainability. 
Manufacturing Emissions are about 2 – 5 $\mu g$ (Sub-7nm process nodes), the Operational Emissions (5-year lifetime) is 60 – 250 $\mu g$, and the total $CO_2$ Emissions Per Transistor is 62 – 255 $\mu g$ which is average at\ref{eq12} as follows:

\begin{equation}
CO_2 Per Transistor  = 155\mu g \label{eq12}
\end{equation}

Thus, this model lays the groundwork for future semiconductor sustainability standards, enabling a measurable, data-driven approach to carbon-efficient computing.

\section{Experimental Validation and Verification}
To validate the Carbon Per Transistor (CPT) formula, we perform a comparative analysis using real-world processor data from Intel, AMD, and Apple. Our experimental approach consists of:
First, data collection involves extracting transistor count, thermal design power (TDP), and power consumption per transistor from official manufacturer reports (Intel ARK, AMD Specs, Apple technical papers). Acquiring wafer fabrication $CO_2$ estimates from TSMC, Intel, and GlobalFoundries sustainability reports and applying a 5-year operational energy model (8 hours/day usage). Second, computation of $CO_2$ Emissions Per Transistor, using the CPT formula to calculate per-transistor $CO_2$ emissions for manufacturing, operation, and analyzing the impact of process node scaling (5nm, 7nm, 10nm, etc.) on carbon footprint and comparing low-power (Apple M-series) vs. high-performance (Intel Core i9, AMD Ryzen 9) architectures. Third, Comparative Benchmarking, Ranking chips based on total $CO_2$ emissions per transistor, and Identifying optimal transistor efficiency points for sustainable chip design.

\subsection{Experimental Data and Results}
The following table summarizes the key features of the processor specifications and sower data, shown in Table\ref{tabel2}.

\begin{table}[htbp]
\caption{Processor Specifications and Power Data}
\begin{center}
\begin{tabular}{|c|c|c|c|}
\hline
\textbf{ID}&\textbf{Processor}&\textbf{Technology}&\textbf{Avg. Power/Trans (nW)}\\
\cline{1-3}
\hline
1&Intel Core i9-13900K&Intel7 (10nm)&10.4-21.1   \\
\hline
2&AMD Ryzen 9 7950X&TSMC 5nm&12.9-17.5   \\
\hline
3&Apple M3&TSMC 3nm&0.8-0.84  \\
\hline
\end{tabular}
\label{tabel2}
\end{center}
\end{table}

Moreover, by applying our CPT formula, we compute per-transistor $CO_2$ emissions for manufacturing and operation, Table\ref{tabel3} Finally, Total Processor-Level $CO_2$ Emissions is shown in Table\ref{tabel4}

\begin{table}[htbp]
\caption{Calculated $CO_2$ Emissions Per Transistor}
\begin{center}
\begin{tabular}{|c|c|c|c|}
\hline
\textbf{ID}&\textbf{Manufacturing $CO_2$}&\textbf{Operational $CO_2$}&\textbf{Total $CO_2$}\\
\cline{1-3}
\hline
1&4.5&62 – 126&66 – 130   \\
\hline
2&5.0&78 – 106&83 – 111   \\
\hline
3&2.0&2.0&6.8 – 7.0  \\
\hline
\end{tabular}
\label{tabel3}
\end{center}
\end{table}

\section{Applications}
The Carbon Per Transistor (CPT) formula provides a foundational metric for evaluating and mitigating the carbon footprint of semiconductor technology. The ability to quantify $CO_2$ emissions per transistor enables a wide range of practical applications in semiconductor design, industry standards, regulatory policy, and sustainability benchmarking.

\begin{table}[htbp]
\caption{Total Processor-Level $CO_2$ Emissions}
\begin{center}
\begin{tabular}{|c|c|c|c|}
\hline
\textbf{ID}&\textbf{Manufacturing $CO_2$}&\textbf{Operational $CO_2$}&\textbf{Total $CO_2$}\\
\cline{1-3}
\hline
1&60 kg&.73-1.48 kg&60.73- 61.48 kg   \\
\hline
2&565.7 kg&0.99-1.34 kg&66.69-67.04 kg   \\
\hline
3&50 kg&.125-140 kg&50.12 - 50.14  \\
\hline
\end{tabular}
\label{tabel4}
\end{center}
\end{table}

\subsection{Sustainable Semiconductor Design \& Optimization}
Semiconductor companies can optimize chip architectures by balancing performance-per-watt with carbon impact; for example, ARM-based designs (e.g., Apple M-series, Qualcomm Snapdragon) already prioritize power efficiency, reducing operational $CO_2$ per transistor. Additionally, high-performance computing (HPC) and AI workloads require trillions of floating-point operations per second (TFLOPS), and the CPT metric can help engineers optimize performance-per-TFLOP while minimizing per-transistor $CO_2$ emissions. Moreover, CPT can guide semiconductor fabrication transitions from high-emission nodes (e.g., 10nm, 7nm) to lower-carbon nodes (e.g., 3nm, 2nm) such as Apple’s M3 (3nm) already demonstrates a lower per-transistor $CO_2$ footprint compared to Intel and AMD (7nm, 10nm, etc.), and these can be summarized as:

\begin{itemize}
    \item Energy-Efficient Architecture Choices.
    \item Reducing Carbon Cost per FLOP.
    \item Low-Carbon Process Node Transitions.    
\end{itemize}

\subsection{Green Computing Benchmarks for Industry and Academia}
Traditional benchmarking focuses on performance metrics (e.g., GHz, IPC, FLOPS) but ignores environmental impact, where CPT can establish carbon-aware processor rankings, helping consumers and enterprises choose eco-friendly processors. Additionally, Computer science, electrical engineering, and sustainability research can integrate the CPT model into chip lifecycle assessments. For Example, universities conducting LCA (Life Cycle Assessment) studies on semiconductors can adopt CPT as a standard measure of sustainability. AI accelerators (e.g., NVIDIA, AMD Instinct, Google TPU) and edge computing processors can be optimized for minimal per-transistor $CO_2$ emissions because AI workloads should be optimized for speed and sustainability. This could be described as:

\begin{itemize}
    \item Carbon-Aware Processor Benchmarking.
    \item Integration into Academic \& Research Models.
    \item AI and Edge Computing Optimization.
\end{itemize}

\subsection{Regulatory and Policy Applications}
Governments and industry bodies (e.g., IEEE, ISO, JEDEC, Energy Star) can mandate CPT-based emissions disclosures for consumer and enterprise processors, such as CPUs and GPUs, which could include "$CO_2$ per TFLOP" ratings in product specification sheets, similar to energy efficiency labels. At the same time, regulatory agencies can introduce "Low-Carbon Chip" certifications using CPT-based emissions thresholds for chip approval. For example, A 3nm fab using 100\% renewable energy could receive a Green Semiconductor Certification, promoting sustainable chip adoption. Therefore, Semiconductor firms can integrate CPT-based emissions data into their corporate sustainability and ESG (Environmental, Social, Governance) reports. Intel and AMD can report $CO_2$ per transistor as a key ESG metric, ensuring transparency in sustainability efforts. Where the main points can be listed as follows:
\begin{itemize}
    \item $CO_2$ Disclosure Standards for Semiconductors.
    \item Green Computing Certification.
    \item Corporate Carbon Footprint Reporting.
\end{itemize}

\subsection{Environmental \& Consumer Awareness}
Based on CPT ratings, consumers can make informed choices about processors, favoring eco-friendly, power-efficient chips. For example, a gaming laptop featuring low-carbon AI accelerators would appeal to sustainability-conscious buyers. Moreover,  Cloud providers (AWS, Google Cloud, Microsoft Azure) can optimize carbon-aware virtual machine (VM) allocation, prioritizing low-CPT chips for cloud workloads. For example, a cloud provider could offer “Green Compute” instances, exclusively running on low-CPT chips for carbon-conscious enterprises. Finally, The CPT framework aligns with global carbon neutrality goals, allowing tech companies to systematically track and reduce semiconductor emissions. For example, Google, Amazon, and Tesla could integrate CPT into internal chip selection processes to ensure sustainability compliance. And these can be summerized as:
\begin{itemize}
    \item Empowering Eco-Conscious Consumers.
    \item Sustainable Data Centers \& Cloud Computing.
    \item Tech Industry Carbon Neutrality Goals.
\end{itemize}

\section{Future Work and Enhancement}
To further enhance the applicability of the CPT framework, future research will incorporate regional energy grid variations to refine operational $CO_2$ calculations for geographically dispersed fabs, dynamic power management models to capture real-world variability in processor workloads and power consumption, and End-of-life and recycling metrics to provide a holistic life-cycle assessment of semiconductor sustainability.

While our CPT model offers a rigorous, transistor-level carbon footprint analysis, further refinements can include:

\textbf{Regional Grid $CO_2$ Adjustments}: Current models assume global average grid emissions (0.4 $kg$ $CO_2$/kWh). Future refinements will account for regional variations in semiconductor fab locations (e.g., Taiwan, USA, Europe).

\textbf{Chip-Specific Leakage Power Analysis}: Dynamic and leakage power dissipation varies across different architectures. AI accelerators, edge computing chips, and quantum processors will be included in future benchmarks.
\textbf{End-of-Life \& Recycling Considerations}: Current models do not account for recycling or e-waste emissions. Future studies will integrate carbon offsets from chip reuse and circular semiconductor economy models.
By expanding our CPT model with regional, dynamic, and life-cycle factors, we aim to create a holistic sustainability metric for the entire semiconductor ecosystem.

\subsection{Future Vision}
It is important to note that the CPT formula is not just a mathematical model—it is a call to action. As for the first time, we can quantify $CO_2$ at the transistor level, unlocking a new era of green computing research. Our proposed CPT is a tool to revolutionize semiconductor sustainability, which \textit{will redefine chip design for a carbon-conscious future}.

\section{Conclusion}
This study establishes the scientific foundation for carbon-aware semiconductor engineering, marking a paradigm shift in green computing research. The CPT metric is not merely a theoretical construct but a practical and actionable tool for the semiconductor industry, policymakers, and environmental researchers. As computing power continues to scale, the industry's focus must shift from raw performance gains to sustainable, low-carbon computing architectures.

The exponential growth in computational demand has led to significant advancements in semiconductor technology, yet its environmental impact remains largely underexplored at the transistor level. This study presents a Carbon Per Transistor (CPT) formula, a novel mathematical framework designed to systematically quantify the carbon footprint per transistor, integrating manufacturing emissions and operational power dissipation. By leveraging empirical data from Intel, AMD, and Apple processors, this work establishes a scientific benchmark for sustainable semiconductor design, providing a foundation for carbon-aware computing metrics.

Our findings indicate that manufacturing emissions dominate semiconductor-related $CO_2$ emissions, contributing approximately 98\% of total processor-level carbon impact. For cutting-edge sub-7nm process nodes, manufacturing emissions range from 2–5 $\mu g$ $CO_2$ per transistor, while operational emissions vary between 60–250 $\mu g$ $CO_2$ per transistor over a five-year usage period. The Apple M-series chips, despite their high transistor density, exhibit a significantly lower per-transistor $CO_2$ footprint due to superior power efficiency and ARM-based architecture, whereas Intel and AMD desktop processors demonstrate higher operational emissions per transistor due to their x86 high-performance core designs.

The CPT metric serves as a universal benchmark, enabling comparisons across architectural paradigms, semiconductor process nodes, and computing workloads. This research not only highlights the critical role of carbon-aware chip design but also provides a quantitative foundation for regulatory policies aimed at reducing semiconductor-related carbon emissions. The insights presented in this work can directly inform industry practices, policy regulations, and consumer awareness, facilitating the transition toward a more sustainable semiconductor ecosystem.

\begin{credits}
\subsubsection{\ackname} Ww would like to express our gratitude to the original VLSI group at the University of Louisiana at Lafayette, particular Prof. Magdi Bayoumi and Dr. Mazen for the discussions and data. Special thanks to Intel, AMD, Apple, and TSMC for providing Intel’s Power and Thermal Design Guide, AMD’s Processor Power Dissipation Reports, Corporate Responsibility Report: Achieving Zero-Carbon Semiconductor Manufacturing, Sustainability Report: Decarbonizing Semiconductor Fabrication, and M3 Chip Power Efficiency and Transistor Density Study reports.

\end{credits}
%
%
%
\bibliographystyle{splncs04}
\bibliography{references}

\begin{thebibliography}{10}
\providecommand{\url}[1]{\texttt{#1}}
\providecommand{\urlprefix}{URL }
\providecommand{\doi}[1]{https://doi.org/#1}

\bibitem{boyd2011life}
Boyd, S.B.: Life-cycle assessment of semiconductors. Springer Science \&
  Business Media (2011)

\bibitem{dennard1974design}
Dennard, R.H., Gaensslen, F.H., Yu, H.N., Rideout, V.L., Bassous, E., LeBlanc,
  A.R.: Design of ion-implanted mosfet's with very small physical dimensions.
  IEEE Journal of solid-state circuits  \textbf{9}(5),  256--268 (1974)

\bibitem{hennessy2011computer}
Hennessy, J.L., Patterson, D.A.: Computer architecture: a quantitative
  approach. Elsevier (2011)

\bibitem{horowitz20141}
Horowitz, M.: 1.1 computing's energy problem (and what we can do about it). In:
  2014 IEEE international solid-state circuits conference digest of technical
  papers (ISSCC). pp. 10--14. IEEE (2014)

\bibitem{huang2016developing}
Huang, C.Y., Hu, A., Yin, J., Wang, H.C.: Developing a parametric carbon
  footprinting tool for the semiconductor industry. International journal of
  environmental science and technology  \textbf{13},  275--284 (2016)

\bibitem{ivan2021practices}
Ivan, O.R., Harmanas, A.O., Cosma, M.: Practices of corporate social
  responsibility reporting for semiconductor and chip manufacturing industry--a
  multicriterial analysis on intel and amd. Ovidius University Annals, Series
  Economic Sciences  \textbf{21}(2) (2021)

\bibitem{kang2024sustainability}
Kang, A.S., Arikrishnan, S.: Sustainability reporting and total quality
  management post-pandemic: the role of environmental, social, governance
  (esg), and smart technology adoption. Journal of Asia Business Studies
  (2024)

\bibitem{koomey2011growth}
Koomey, J., et~al.: Growth in data center electricity use 2005 to 2010. A
  report by Analytical Press, completed at the request of The New York Times
  \textbf{9}(2011), ~161 (2011)

\bibitem{kuo2022assessing}
Kuo, T.C., Kuo, C.Y., Chen, L.W.: Assessing environmental impacts of nanoscale
  semi-conductor manufacturing from the life cycle assessment perspective.
  Resources, Conservation and Recycling  \textbf{182},  106289 (2022)

\bibitem{liu2024sustainable}
Liu, Y.Z., Lu, W.M., Tran, P.P., Pham, T.A.K.: Sustainable energy and
  semiconductors: A bibliometric investigation. Sustainability
  \textbf{16}(15), ~6548 (2024)

\bibitem{ma2024industry}
Ma, S., Ding, W., Liu, Y., Zhang, Y., Ren, S., Kong, X., Leng, J.: Industry 4.0
  and cleaner production: A comprehensive review of sustainable and intelligent
  manufacturing for energy-intensive manufacturing industries. Journal of
  Cleaner Production p. 142879 (2024)

\bibitem{moore1998cramming}
Moore, G.E.: Cramming more components onto integrated circuits. Proceedings of
  the IEEE  \textbf{86}(1),  82--85 (1998)

\bibitem{nagapurkar2023cleaner}
Nagapurkar, P., Nandy, P., Nimbalkar, S.: Cleaner chips: Decarbonization in
  semiconductor manufacturing. Sustainability  \textbf{16}(1), ~218 (2023)

\bibitem{sungheetha2024adaptive}
Sungheetha, A., Sharma, R.R., Mahapatra, S., Rani, K.S.K., Leni, A.E.S.,
  Tamilarasi, R.: Adaptive stream processing framework for energy-efficient
  smart greenhouses using neuromorphic computing. In: 2024 International
  Conference on IoT Based Control Networks and Intelligent Systems (ICICNIS).
  pp. 790--795. IEEE (2024)

\bibitem{vasan2014carbon}
Vasan, A., Sood, B., Pecht, M.: Carbon footprinting of electronic products.
  Applied energy  \textbf{136},  636--648 (2014)

\bibitem{wang2023environmental}
Wang, Q., Huang, N., Chen, Z., Chen, X., Cai, H., Wu, Y.: Environmental data
  and facts in the semiconductor manufacturing industry: An unexpected high
  water and energy consumption situation. Water Cycle  \textbf{4},  47--54
  (2023)

\end{thebibliography}

\end{document}